\input harvmac
\input epsf
\input amssym
%
%
\noblackbox
\newcount\figno
\figno=0
\def\fig#1#2#3{
\par\begingroup\parindent=0pt\leftskip=1cm\rightskip=1cm\parindent=0pt
\baselineskip=11pt
\global\advance\figno by 1
\midinsert
\epsfxsize=#3
\centerline{\epsfbox{#2}}
\vskip -21pt
{\bf Fig.\ \the\figno: } #1\par
\endinsert\endgroup\par
}
\def\figlabel#1{\xdef#1{\the\figno}}
\def\encadremath#1{\vbox{\hrule\hbox{\vrule\kern8pt\vbox{\kern8pt
\hbox{$\displaystyle #1$}\kern8pt}
\kern8pt\vrule}\hrule}}

\def\frac#1#2{{#1 \over #2}}

\def\p{\partial}
\def\semi{\subset\kern-1em\times\;}
\def\bar#1{\overline{#1}}
\def\sqr#1#2{{\vcenter{\vbox{\hrule height.#2pt
\hbox{\vrule width.#2pt height#1pt \kern#1pt \vrule width.#2pt}
\hrule height.#2pt}}}}

\def\p{\partial}

\def\ad{\bar a}

\def\p{\partial}

\def\ct{\tilde{c}}

\def\la{\ell_{Ads}}
\def\ls{\ell_{S^p} }
\def\lab{\overline{\ell}_{Ads}}
\def\lsb{\overline{\ell}_{S^p} }
\def\taub{\overline{\tau}}
\def\etam{\eta_{{\rm max}}}
\def\Lt{\tilde{L}}
\def\ltwo{\ell_{S^2}}
\def\Xb{\overline{X}}

%

%


\lref\fareytale{J.~M.~Maldacena and A.~Strominger,
  ``AdS(3) black holes and a stringy exclusion principle,''
  JHEP {\bf 9812}, 005 (1998)
  [arXiv:hep-th/9804085];
  R.~Dijkgraaf, J.~M.~Maldacena, G.~W.~Moore and E.~Verlinde,
  ``A black hole farey tail,''
  arXiv:hep-th/0005003.

}

\lref\generalRRRR{
  M.~B.~Green and J.~H.~Schwarz,
  ``Supersymmetrical Dual String Theory. 2. Vertices And Trees,''
  Nucl.\ Phys.\ B {\bf 198}, 252 (1982);
  D.~J.~Gross and E.~Witten,
  ``Superstring Modifications Of Einstein's Equations,''
  Nucl.\ Phys.\ B {\bf 277}, 1 (1986);
  W.~Lerche, B.~E.~W.~Nilsson and A.~N.~Schellekens,
  ``Heterotic String Loop Calculation Of The Anomaly Cancelling Term,''
  Nucl.\ Phys.\ B {\bf 289}, 609 (1987);
  M.~J.~Duff, J.~T.~Liu and R.~Minasian,
  ``Eleven-dimensional origin of string / string duality: A one-loop test,''
  Nucl.\ Phys.\ B {\bf 452}, 261 (1995)
  [arXiv:hep-th/9506126];
   M.~B.~Green, M.~Gutperle and P.~Vanhove,
  ``One loop in eleven dimensions,''
  Phys.\ Lett.\ B {\bf 409}, 177 (1997)
  [arXiv:hep-th/9706175];
    J.~G.~Russo and A.~A.~Tseytlin,
  ``One-loop four-graviton amplitude in eleven-dimensional supergravity,''
  Nucl.\ Phys.\ B {\bf 508}, 245 (1997)
  [arXiv:hep-th/9707134]; P.~S.~Howe and D.~Tsimpis,
  ``On higher-order corrections in M theory,''
  JHEP {\bf 0309}, 038 (2003)
  [arXiv:hep-th/0305129].
}

\lref\masakiref{
  N.~Iizuka and M.~Shigemori,
  ``A Note on D1-D5-J System and 5D Small Black Ring,''
  arXiv:hep-th/0506215.
}

\lref\russuss{  J.~G.~Russo and L.~Susskind,
  ``Asymptotic level density in heterotic string theory and rotating black
  holes,''
  Nucl.\ Phys.\ B {\bf 437}, 611 (1995)
  [arXiv:hep-th/9405117].
  }

\lref\KLatt{
  P.~Kraus and F.~Larsen,
  ``Attractors and black rings,''
  arXiv:hep-th/0503219.
}

\lref\supertube{D.~Mateos and P.~K.~Townsend,
  ``Supertubes,''
  Phys.\ Rev.\ Lett.\  {\bf 87}, 011602 (2001)
  [arXiv:hep-th/0103030];
  R.~Emparan, D.~Mateos and P.~K.~Townsend,
  ``Supergravity supertubes,''
  JHEP {\bf 0107}, 011 (2001)
  [arXiv:hep-th/0106012].
}

\lref\Mathurrev{
S.~D.~Mathur,``The fuzzball proposal for black holes: An elementary review,''
arXiv:hep-th/0502050.
}

\lref\StromBTZ{
A.~Strominger,
 ``Black hole entropy from near-horizon microstates'',
JHEP {\bf 9802}, 009 (1998);
[arXiv:hep-th/9712251]; V.~Balasubramanian and F.~Larsen,
``Near horizon geometry and black holes in four dimensions'',
Nucl.\ Phys.\ B {\bf 528}, 229 (1998);
[arXiv:hep-th/9802198].
}

\lref\MSW{
J.~M.~Maldacena, A.~Strominger and E.~Witten,
``Black hole entropy in M-theory'',
JHEP {\bf 9712}, 002 (1997);
[arXiv:hep-th/9711053].
}

\lref\HMM{
J.~A.~Harvey, R.~Minasian and G.~W.~Moore,
``Non-abelian tensor-multiplet anomalies,''
 JHEP {\bf 9809}, 004 (1998)
  [arXiv:hep-th/9808060].
}

\lref\tseytRRRR{  A.~A.~Tseytlin,
  ``R**4 terms in 11 dimensions and conformal anomaly of (2,0) theory,''
  Nucl.\ Phys.\ B {\bf 584}, 233 (2000)
  [arXiv:hep-th/0005072].
}

\lref\antRRRR{  I.~Antoniadis, S.~Ferrara, R.~Minasian and K.~S.~Narain,
  ``R**4 couplings in M- and type II theories on Calabi-Yau spaces,''
  Nucl.\ Phys.\ B {\bf 507}, 571 (1997)
  [arXiv:hep-th/9707013].
 }

\lref\WittenMfive{ E.~Witten,
  ``Five-brane effective action in M-theory,''
  J.\ Geom.\ Phys.\  {\bf 22}, 103 (1997)
  [arXiv:hep-th/9610234].
}

\lref\wittenAdS{ E.~Witten,
  ``Anti-de Sitter space and holography,''
  Adv.\ Theor.\ Math.\ Phys.\  {\bf 2}, 253 (1998)
  [arXiv:hep-th/9802150].
  }

\lref\iosef{  I.~Bena,
  ``Splitting hairs of the three charge black hole,''
  Phys.\ Rev.\ D {\bf 70}, 105018 (2004)
  [arXiv:hep-th/0404073].
}

\lref\brownhen{  J.~D.~Brown and M.~Henneaux,
 ``Central Charges In The Canonical Realization Of Asymptotic Symmetries: An
  Example From Three-Dimensional Gravity,''
  Commun.\ Math.\ Phys.\  {\bf 104}, 207 (1986).
  }

\lref\wald{
  R.~M.~Wald,
  ``Black hole entropy is the Noether charge,''
  Phys.\ Rev.\ D {\bf 48}, 3427 (1993)
  [arXiv:gr-qc/9307038].
R.~Wald,
Phys.\ Rev.\ D {\bf 48} R3427 (1993);
   V.~Iyer and R.~M.~Wald,
  ``Some properties of Noether charge and a proposal for dynamical black hole
  entropy,''
  Phys.\ Rev.\ D {\bf 50}, 846 (1994)
  [arXiv:gr-qc/9403028].
 ``A Comparison of Noether charge and Euclidean methods for computing the
  entropy of stationary black holes,''
  Phys.\ Rev.\ D {\bf 52}, 4430 (1995)
  [arXiv:gr-qc/9503052].
}

\lref\senrescaled{  A.~Sen,
  ``How does a fundamental string stretch its horizon?,''
  JHEP {\bf 0505}, 059 (2005)
  [arXiv:hep-th/0411255];
   ``Black holes, elementary strings and holomorphic anomaly,''
  arXiv:hep-th/0502126.
   ``Stretching the horizon of a higher dimensional small black hole,''
  arXiv:hep-th/0505122.
  }

\lref\saidasoda{
  H.~Saida and J.~Soda,
  ``Statistical entropy of BTZ black hole in higher curvature gravity,''
  Phys.\ Lett.\ B {\bf 471}, 358 (2000)
  [arXiv:gr-qc/9909061].
}

\lref\attract{
S.~Ferrara, R.~Kallosh and A.~Strominger,
``N=2 extremal black holes'',
Phys.\ Rev.\ D {\bf 52}, 5412 (1995),
[arXiv:hep-th/9508072];
 A.~Strominger,
 ``Macroscopic Entropy of $N=2$ Extremal Black Holes'',
 Phys.\ Lett.\ B {\bf 383}, 39 (1996),
[arXiv:hep-th/9602111];
S.~Ferrara and R.~Kallosh,
``Supersymmetry and Attractors'',
Phys.\ Rev.\ D {\bf 54}, 1514 (1996),
[arXiv:hep-th/9602136];
``Universality of Supersymmetric Attractors'',
Phys.\ Rev.\ D {\bf 54}, 1525 (1996),
[arXiv:hep-th/9603090];
R.~Kallosh, A.~Rajaraman and W.~K.~Wong,
``Supersymmetric rotating black holes and attractors'',
Phys.\ Rev.\ D {\bf 55}, 3246 (1997),
[arXiv:hep-th/9611094];
A~Chou, R.~Kallosh, J.~Rahmfeld, S.~J.~Rey, M.~Shmakova and W.~K.~Wong,
``Critical points and phase transitions in 5d compactifications of M-theory''.
Nucl.\ Phys.\ B {\bf 508}, 147 (1997);
[arXiv:hep-th/9704142].
}

\lref\moore{G.~W.~Moore,``Attractors and arithmetic'',
[arXiv:hep-th/9807056];
``Arithmetic and attractors'',
[arXiv:hep-th/9807087];
``Les Houches lectures on strings and arithmetic'',
[arXiv:hep-th/0401049];
B.~R.~Greene and C.~I.~Lazaroiu,
``Collapsing D-branes in Calabi-Yau moduli space. I'',
Nucl.\ Phys.\ B {\bf 604}, 181 (2001),
[arXiv:hep-th/0001025].
}

\lref\ChamseddinePI{
  A.~H.~Chamseddine, S.~Ferrara, G.~W.~Gibbons and R.~Kallosh,
  ``Enhancement of supersymmetry near 5d black hole horizon,''
  Phys.\ Rev.\ D {\bf 55}, 3647 (1997)
  [arXiv:hep-th/9610155].
}

\lref\denef{  
F.~Denef,``Supergravity flows and D-brane stability'',
JHEP {\bf 0008}, 050 (2000), [arXiv:hep-th/0005049];
``On the correspondence between D-branes and stationary supergravity
 solutions of type II Calabi-Yau compactifications'',
[arXiv:hep-th/0010222];
``(Dis)assembling special Lagrangians'',
[arXiv:hep-th/0107152].
  B.~Bates and F.~Denef,
   ``Exact solutions for supersymmetric stationary black hole composites,''
  arXiv:hep-th/0304094.
}

\lref\OSV{H.~Ooguri, A.~Strominger and C.~Vafa,
``Black hole attractors and the topological string'',
Phys.\ Rev.\ D {\bf 70}, 106007 (2004),
[arXiv:hep-th/0405146];
E.~Verlinde,
``Attractors and the holomorphic anomaly'',
[arXiv:hep-th/0412139];
  }

 \lref\DDMP{
A.~Dabholkar, F.~Denef, G.~W.~Moore and B.~Pioline,
``Exact and asymptotic degeneracies of small black holes'',
[arXiv:hep-th/0502157].
}

\lref\curvcorr{A.~Dabholkar,
``Exact counting of black hole microstates",
[arXiv:hep-th/0409148],
A.~Dabholkar, R.~Kallosh and A.~Maloney,
``A stringy cloak for a classical singularity'',
JHEP {\bf 0412}, 059 (2004),
[arXiv:hep-th/0410076].
}
\lref\bkmicro{
 I.~Bena and P.~Kraus,
 ``Microscopic description of black rings in AdS/CFT'',
JHEP {\bf 0412}, 070 (2004)
  [arXiv:hep-th/0408186].
}
\lref\cgms{
M.~Cyrier, M.~Guica, D.~Mateos and A.~Strominger,
``Microscopic entropy of the black ring'',
[arXiv:hep-th/0411187].
}

\lref\GunaydinBI{
  M.~Gunaydin, G.~Sierra and P.~K.~Townsend,
  ``The Geometry Of N=2 Maxwell-Einstein Supergravity And Jordan
  Algebras'',
  Nucl.\ Phys.\ B {\bf 242}, 244 (1984);
  ``Gauging The D = 5 Maxwell-Einstein Supergravity Theories:
   More On Jordan Algebras,''
  Nucl.\ Phys.\ B {\bf 253}, 573 (1985).
}

\lref\deWitCR{
  B.~de Wit and A.~Van Proeyen,
  ``Broken sigma model isometries in very special geometry,''
  Phys.\ Lett.\ B {\bf 293}, 94 (1992)
  [arXiv:hep-th/9207091].
}

\lref\CadavidBK{
  A.~C.~Cadavid, A.~Ceresole, R.~D'Auria and S.~Ferrara,
  ``Eleven-dimensional supergravity compactified on Calabi-Yau threefolds,''
  Phys.\ Lett.\ B {\bf 357}, 76 (1995)
  [arXiv:hep-th/9506144].
}

\lref\PapadopoulosDA{
  G.~Papadopoulos and P.~K.~Townsend,
  ``Compactification of D = 11 supergravity on spaces of exceptional
  holonomy,''
  Phys.\ Lett.\ B {\bf 357}, 300 (1995)
  [arXiv:hep-th/9506150].
}

\lref\AntoniadisCY{
  I.~Antoniadis, S.~Ferrara and T.~R.~Taylor,
  ``N=2 Heterotic Superstring and its Dual Theory in Five Dimensions,''
  Nucl.\ Phys.\ B {\bf 460}, 489 (1996)
  [arXiv:hep-th/9511108].
}

\lref\GauntlettNW{
  J.~P.~Gauntlett, J.~B.~Gutowski, C.~M.~Hull, S.~Pakis and H.~S.~Reall,
  ``All supersymmetric solutions of minimal supergravity in five dimensions,''
  Class.\ Quant.\ Grav.\  {\bf 20}, 4587 (2003)
  [arXiv:hep-th/0209114]}

\lref\gutow{
  J.~B.~Gutowski and H.~S.~Reall,
  ``General supersymmetric AdS(5) black holes'',
  JHEP {\bf 0404}, 048 (2004)
  [arXiv:hep-th/0401129];
  J.~B.~Gutowski,
  ``Uniqueness of five-dimensional supersymmetric black holes'',
  JHEP {\bf 0408}, 049 (2004)
  [arXiv:hep-th/0404079].
}

\lref\BenaDE{
  I.~Bena and N.~P.~Warner,
  ``One ring to rule them all ... and in the darkness bind them?'',
  [arXiv:hep-th/0408106].
}
\lref\BMPV{ J.~C.~Breckenridge, R.~C.~Myers, A.~W.~Peet and C.~Vafa,
``D-branes and spinning black holes'',
Phys.\ Lett.\ B {\bf 391}, 93 (1997);
[arXiv:hep-th/9602065].
}

\lref\EEMR{H.~Elvang, R.~Emparan, D.~Mateos and H.~S.~Reall,
``Supersymmetric black rings and three-charge supertubes'',
  Phys.\ Rev.\ D {\bf 71}, 024033 (2005);
  [arXiv:hep-th/0408120].
}

\lref\ElvangRT{
  H.~Elvang, R.~Emparan, D.~Mateos and H.~S.~Reall,
  ``A supersymmetric black ring'',
  Phys.\ Rev.\ Lett.\  {\bf 93}, 211302 (2004)
  [arXiv:hep-th/0407065].
}

\lref\BenaWT{
  I.~Bena and P.~Kraus,
  ``Three charge supertubes and black hole hair,''
  Phys.\ Rev.\ D {\bf 70}, 046003 (2004)
  [arXiv:hep-th/0402144].
}

\lref\GauntlettQY{ J.~P.~Gauntlett and J.~B.~Gutowski, ``General
Concentric  Black Rings'', [arXiv:hep-th/0408122]. J.~P.~Gauntlett
and J.~B.~Gutowski, ``Concentric  black rings'',
[arXiv:hep-th/0408010].
}

\lref\denefa{
  F.~Denef,
   ``Supergravity flows and D-brane stability'',
JHEP {\bf 0008}, 050 (2000), [arXiv:hep-th/0005049].
}

\lref\denefc{
  B.~Bates and F.~Denef,
``Exact solutions for supersymmetric stationary black hole composites'',
[arXiv:hep-th/0304094].
}

\lref\SenPU{
A.~Sen,
``Black holes, elementary strings and holomorphic anomaly'',
 [arXiv:hep-th/0502126].
}

\lref\CardosoFP{
  K.~Behrndt, G.~Lopes Cardoso, B.~de Wit, D.~Lust, T.~Mohaupt and W.~A.~Sabra,
  ``Higher-order black-hole solutions in N = 2 supergravity and Calabi-Yau
  string backgrounds,''
  Phys.\ Lett.\ B {\bf 429}, 289 (1998)
  [arXiv:hep-th/9801081];
G.~L.~Cardoso, B.~de Wit, J.~Kappeli and T.~Mohaupt,
``Examples of stationary BPS solutions in N = 2 supergravity theories  with
$R^2$-interactions,''
Fortsch.\ Phys.\  {\bf 49}, 557 (2001)
[arXiv:hep-th/0012232];
``Stationary BPS solutions in N = 2 supergravity with $R^2 $ interactions'',
JHEP {\bf 0012}, 019 (2000)
[arXiv:hep-th/0009234];
  ``Supersymmetric black hole solutions with $R^2$ interactions'',
[arXiv:hep-th/0003157];
  G.~Lopes Cardoso, B.~de Wit and T.~Mohaupt,
  ``Area law corrections from state counting and supergravity'',
  Class.\ Quant.\ Grav.\  {\bf 17}, 1007 (2000)
  [arXiv:hep-th/9910179];
  ``Macroscopic entropy formulae and non-holomorphic corrections for
  supersymmetric black holes'',
  Nucl.\ Phys.\ B {\bf 567}, 87 (2000)
  [arXiv:hep-th/9906094];
  ``Deviations from the area law for supersymmetric black holes'',
  Fortsch.\ Phys.\  {\bf 48}, 49 (2000)
  [arXiv:hep-th/9904005];
  ``Corrections to macroscopic supersymmetric black-hole entropy'',
  Phys.\ Lett.\ B {\bf 451}, 309 (1999)
  [arXiv:hep-th/9812082].
}

\lref\BenaTD{
  I.~Bena, C.~W.~Wang and N.~P.~Warner,
  ``Black rings with varying charge density'',
[arXiv:hep-th/0411072].
}

\lref\BenaWV{
I.~Bena, ``Splitting hairs of the three charge black hole'',
Phys.\ Rev.\ D {\bf 70}, 105018 (2004),
[arXiv:hep-th/0404073].
}

\lref\rings{ H.~Elvang, R.~Emparan, D.~Mateos and H.~S.~Reall, ``A
supersymmetric black ring,''  Phys. Rev. Lett.  {\bf 93}, 211302
(2004) [arXiv:hep-th/0407065]; ``Supersymmetric black rings and
three-charge supertubes,'' Phys. Rev. D {\bf 71}, 024033 (2005)
[arXiv:hep-th/0408120]; I.~Bena and N.~P.~Warner, ``One ring to
rule them all ... and in the darkness bind them?,''
arXiv:hep-th/0408106; J.~P.~Gauntlett and J.~B.~Gutowski,
``General concentric black rings,'' Phys.\ Rev. D {\bf 71}, 045002
(2005) [arXiv:hep-th/0408122].}

\lref\hensken{  M.~Henningson and K.~Skenderis,
  ``The holographic Weyl anomaly,''
  JHEP {\bf 9807}, 023 (1998)
  [arXiv:hep-th/9806087].
  }

\lref\balkraus{  V.~Balasubramanian and P.~Kraus,
  ``A stress tensor for anti-de Sitter gravity,''
  Commun.\ Math.\ Phys.\  {\bf 208}, 413 (1999)
  [arXiv:hep-th/9902121].
  }

\lref\MarolfFY{
  D.~Marolf and B.~C.~Palmer,
  ``Gyrating strings: A new instability of black strings?,''
  Phys.\ Rev.\ D {\bf 70}, 084045 (2004)
  [arXiv:hep-th/0404139].
}


\Title{\vbox{\baselineskip12pt
}}
{\vbox{\centerline
{Microscopic Black Hole Entropy}\medskip\vbox{\centerline {in Theories with Higher Derivatives}}} } \centerline{Per Kraus\foot{pkraus@physics.ucla.edu} and
Finn Larsen\foot{larsenf@umich.edu}}

\bigskip
\centerline{${}^1$\it{Department of Physics and Astronomy,
UCLA,}}\centerline{\it{ Los Angeles, CA 90095-1547,
USA.}}\vskip.2cm \centerline{${}^2$\it{Michigan Center for
Theoretical Physics, Department of Physics}} \centerline{\it{University of Michigan, Ann
Arbor, MI 48109-1120, USA.}}

\baselineskip15pt

\vskip .3in

\centerline{\bf Abstract}

We discuss higher derivative corrections to black hole entropy in
theories that allow a near horizon $AdS_3\times X$ geometry. In
arbitrary theories with diffeomorphism invariance we show how to
obtain the spacetime central charge in a simple way.   Black hole
entropy then follows from the Euclidean partition function, and we
show that this gives agreement  with Wald's formula. In string
theory there are certain diffeomorphism anomalies that we exploit.
We thereby reproduce some recent computations of corrected entropy
formulas, and extend them to the nonextremal, nonsupersymmetric
context.  Examples include black holes in M-theory on $K3 \times
T^2$, whose entropy  reproduces that of the perturbative heterotic
string with both right and left movers excited and angular
momentum included.  Our anomaly based approach also sheds light on
why exact results have been obtained in four dimensions while
ignoring $R^4$ type corrections.

\Date{June, 2005}
\baselineskip14pt
\newsec{Introduction}
The famous area law of Bekenstein and Hawking relates the entropy
of a black hole to the area of its event horizon as \eqn\BHlaw{ S
= {1\over 4G_D} A_{D-2}~. } In string theory this law has been
verified in examples where the entropy is interpreted
statistically in terms of microstates and the area is that of a
black hole with the same macroscopic charges as the statistical
system. In such computations many details of the string spectrum
are known, implying numerous corrections to the microscopic
theory. Additionally, higher derivative terms in the action modify
the classical geometry and also change the area law \BHlaw\ into
Wald's entropy formula \wald
\eqn\waldlaw{ S = -{1\over 8G_D} \int_{\rm hor} d^{D-2}x\sqrt{h}~
{\delta {\cal L}_D\over\delta R_{\mu\nu\alpha\beta}}
\epsilon^{\mu\nu}\epsilon^{\alpha\beta}~, }
 which takes into
account arbitrary derivative terms in the action. Remarkably,
agreement between microscopics and macroscopics is maintained also
after all these corrections are taken into account, at least in
some examples \refs{\CardosoFP,\MSW,\HMM} .

Recently it was pointed out that there are special cases of this
agreement where the area of  the black hole vanishes at leading
order: $A_{D-2}=0$ \curvcorr. For example, the microstates of the
heterotic string consist of the usual perturbative spectrum. The
black hole with the same classical charges has vanishing area in
the two-derivative approximation to the action, but after higher
derivatives are taken into account the entropy found from
\waldlaw\ agrees with the microscopic result. This example is
important because the microscopics is so simple, which should
facilitate very detailed comparisons that can test the whole
framework and its interpretation. In particular, this seems like
an ideal setting for testing Mathur's conjecture \Mathurrev\ that
all microstates can be realized as distinct geometries.

Ultimately one would like to understand which features of quantum
gravity are responsible for these striking agreements between
radically different representations of black hole physics. The
purpose of this note is to emphasize the central role played by
symmetries,  particularly diffeomorphism invariance and its
anomalies.  Viewed in this light, some of the agreements between
microscopic and macroscopic results seem less surprising.

The key assumption in our approach is the existence of a  near
horizon region that includes an $AdS_3$ factor, even after higher
derivative terms have been included in the  Lagrangian. This
assumption is suggested by the central role played by such near
horizon geometries in the microscopics of black holes with finite
area \StromBTZ. Additionally, in an appropriate duality frame, an
$AdS_3$  factor appears in the corrected geometry in all examples
where derivative corrections have successfully been taken into
account, at least as far as we are aware. The power of the
assumption is that it relates the Lagrangian to the radius of the
$AdS_3$ space and so, {\it via} generalized  Brown-Henneaux
\brownhen\ reasoning, to the central charges $c_{L,R}$ of the
associated conformal field theory. As we will see, the saddlepoint
approximation to the black hole entropy, including all higher
derivative corrections, is then given by the Cardy formula
\eqn\hetex{ S = 2\pi\left[ \sqrt{c_L h_L\over 6} +
\sqrt{c_R h_R\over 6} ~\right] }
where $h_{L,R}$ are the left and
right moving momenta of the near horizon solution. Although the
detailed form of the central charges $c_{L,R}$ depends critically
on the spacetime Lagrangian, it will turn out that the Cardy
formula \hetex\  agrees with Wald's formula \waldlaw\ for general
theories with diffeomorphism invariance.    Thus, computation of
the corrected black hole entropy reduces to finding the central
charges.  We will present a novel method for achieving this ---
 {\it c-extremization} --- which just involves solving algebraic
 equations.   Given a higher derivative Lagrangian it is then
 quite simple to compute the corrected entropy.

Recent work has shown that in favorable cases it is possible to
reproduce  microscopic degeneracies of states to all orders in an
expansion in inverse powers of charges \refs{\senrescaled,\DDMP}.
This result emerges just as naturally in our approach.  Knowledge
of the central charge leads to an expression for the black hole
partition function which, when inverse Laplace transformed as in
\DDMP\ yields the microscopic degeneracies including all power law
corrections.

We stress that our considerations are independent  of spacetime
supersymmetry. This contrasts with the (much) more explicit
approach of \refs{\CardosoFP,\curvcorr} which relies on the full
power of  supergravity. In particular, the usual approach has so
far been limited to four dimensions, where supergravity is best
developed, while our results apply equally in five dimensions.

In string theory there is  additional  structure due to anomalies
which affect diffeomorphism invariance. Some relevant aspects are
discussed in \refs{\MSW,\HMM}. These anomalies ultimately arise
from $M5$-branes on the compactification manifold but they can
also be understood without reference to $M5$-branes, using
standard AdS/CFT reasoning. In this way we recover  formulae from
\refs{\MSW,\HMM} using elementary methods.

A natural context for these considerations is M-theory on
$AdS_3\times S^2\times X$ where $X$ is some Calabi-Yau three-fold.
A particularly striking example arises when $X=K3\times T^2$,  so
that M-theory is dual to heterotic string theory on $T^5$. In this
case we find $c_L=12$ and $c_R=24$ which are indeed the correct
central charges for the heterotic string. The remarkable feature
is that we are sensitive to {\it both} chiral sectors of the
heterotic string, and that we thereby derive the entropy for the
heterotic string with both sectors excited.  This shows that
agreement is possible even without supersymmetry.

The point we wish to emphasize is that the constraints of matching
symmetries and anomalies are enough to explain the successful
entropy comparisons, at least in the cases we have considered. One
puzzle in existing work has been why exact results are obtained by
keeping only $R^2$ corrections, and neglecting higher powers. Here
we see that it is the $R^2$ terms which yield the relevant
diffeomorphism anomalies, and they are uncorrected by additional
higher derivative terms.

The conventional approach of \refs{\CardosoFP,\curvcorr}  involves
near horizon geometries with an $AdS_2$ factor and uses results
from topological string theory \refs{\OSV,\DDMP}. These  $AdS_2$
geometries are related to $AdS_3$ by compactification. The $AdS_3$
perspective is simpler because spacetime symmetries such as the
Virasoro algebra become manifest. On the other hand, in our
approach we have not yet exploited the effects that can be seen
only after compactification. It would be interesting to analyze
how these further constrain the black hole spectrum.

Another open question is to find a  criterion that determines when
a near horizon $AdS_3$ appears from a singular geometry after
derivative corrections are taken into account. This would
characterize any ultimate limitations of our approach.

The remainder of this paper is organized as follows. In section 2
we consider the higher derivative corrections in a rather general
setting, assuming only that the Lagrangian is formed from the
metric in a diffeomorphism invariant way. In section 3 we apply
these considerations to the case of M-theory on $CY_3$. In section 4,
we discuss modifications due to gravitational anomalies and the
appearance of the perturbative heterotic string spectrum. Finally,
in section 5, we conclude with a discussion of how power law
corrections to the entropy are taken into account using our approach.

\newsec{Central charge and black hole entropy}
In this section we first derive an expression for the central
charge in terms of the Lagrangian including higher derivative
corrections. We then review the computation of BTZ black hole
entropy from the central charge. Finally, we combine the results
and write the entropy in a form that agrees with Wald's formula.

\subsec{Computation of the central charge}

We focus on brane configurations that have a near horizon geometry
${\cal M} =$ AdS$_3 \times S^p \times X$, where $X$ is an
arbitrary compact space.  One familiar case is $p=3$, which arises
from the D1-D5 system in IIB string theory, where $X$ is $T^4$ or
$K3$.  This system gives rise to black holes in $D=5$.  Another
important example is $p=2$ which corresponds to $D=4$ black holes made from
wrapping M2-branes and M5-branes on $X=CY_3$.  We will come back
to particular examples later, for now remaining in a general
setting.

We take the near horizon limit, so that we have a theory of (not
necessarily super) gravity on ${\cal M}$.  In this section we will
take the metric to have Euclidean signature.   For our purposes it
is most convenient to perform a Kaluza-Klein reduction on $X$, to
obtain a theory on AdS$_3 \times S^p$.   The action for this
theory is
\eqn\aa{ I = {1 \over 16 \pi G_{p+3} } \int\! d^{p+3}x \sqrt{g}
\,{\cal L}_{p+3}+S_{{\rm bndy}}+S_{CS}~.}
At this stage, ${\cal L}_{p+3}$ is an arbitrary  function of the
gravitational and matter fields, which is diffeomorphism invariant
up to total derivatives that are cancelled by $S_{{\rm bndy}}$. In
particular, it can include arbitrary higher derivative terms. The
boundary terms indicated in \aa\ are needed to have a well-defined
variational principle and also to define the boundary stress
tensor \refs{\hensken,\balkraus}; but we will not need their
explicit form. $S_{CS}$ denotes Chern-Simons terms built out of
gauge fields; we isolate these for reasons that will become
apparent as we proceed.

 We will be assuming that this theory admits
solutions of the form AdS$_3 \times S^p$, over which ${\cal
L}_{p+3}$ takes a constant value. This is indeed the case for the
examples mentioned above. The radii of the two spaces are taken to
be $\la$ and $\ls$.  For a general action there is not necessarily
a single preferred definition of the metric, and so the radii are
defined with respect to some particular choice.

As originally shown by Brown and Henneaux \brownhen,
a theory of gravity on
a space whose noncompact part is   AdS$_3$ corresponds to a
conformal field theory on the two dimensional boundary.  The
conformal field theory has left and right moving central charges,
$c_L$ and $c_R$, which are not necessarily equal.  In this section
we will consider the case in which they {\it are} equal.  This is
true for the D1-D5 system; for the M2-M5-brane case mentioned
above it is only true for the leading  part in an expansion in
charges. We will come back to the case of unequal central charges
later, for now just remarking that it leads to non-diffeomorphism
invariant theories (gravitational anomalies), and so requires
special care.

Our first task is to compute $\la$ and $\ls$.   Suppose we
consider a family of trial solutions with $\la$ and $\ls$ left as
free parameters.  In particular, we write the metric as
\eqn\ab{ds^2 = \la^2\left( d\eta^2 +  \sinh^2 \eta\, d\Omega_2^2
\right)+ \ls^2 d\Omega_p^2~.}
The first two terms give AdS$_3$ in a convenient, but perhaps
slightly unfamiliar, form.
  The actual values of the radii can then be obtained by demanding that
the combination $\la^3 \ls^p{\cal L}_{p+3}$ be stationary under
variation of $\la$ and $\ls$.  Roughly speaking, this can be
thought of as extremizing the bulk action.   A little care is
required to establish that this is the correct procedure.   In
particular, we should recall that when the equations of motion are
satisfied the full action is stationary under variations which
vanish at the boundary; but in our case variations of the radii
lead to variations even at the boundary. Furthermore, we have the
boundary terms in \aa.  A simple way to avoid these complications
is to consider an analytic continuation so that our solutions take
the form $S^3 \times S^p$. Then both complications are absent, and
the total action is clearly proportional to $\la^3 \ls^p{\cal
L}_{p+3}$. Hence this combination must be stationary. Our result
follows after continuation back to AdS$_3 \times S^p$. We note
that in general ${\cal L}_{p+3}$ will be a complicated function of
the radii, incorporating for example the contributions of the
field strengths, whose values are fixed by their equations of
motion.

This discussion  makes it clear why we isolated the Chern-Simon
terms.  These are not necessarily constant over our solution. On
the other hand, being topological they play no role in determining
the radii, or the central charge, so we are free to neglect them
at this stage.

With foresight, we define the central charge function
\eqn\ac{c(\la,\ls) = {3 \Omega_2 \Omega_p \over 32\pi G_{p+3}}
\la^3 \ls^p {\cal L}_{p+3}~,}
and so the actual values of the radii are determined by solving
\eqn\ad{ {\p \over \p \la}c(\la,\ls)|_{\la = \lab} ={\p \over \p
\ls}c(\la,\ls)|_{\ls=\lsb}=0~.}
We wish to establish that $c = c(\lab,\lsb)$ (equal on left and right!) is
indeed the central charge, as defined by the conformal anomaly
\eqn\af{T^i_i = - {c \over 12} ^{(2)}R~,}
of the dual $D=2$ CFT.

To this end, put  the $2D$ CFT on a sphere with metric
\eqn\ag{ds^2 = e^{2\omega}d\Omega_2^2~,}
and focus on the partition function, $Z= e^{-I}$,  as a function
of $\omega$. Under constant shifts of $\omega$ we have
\eqn\ah{ \delta I = {1\over 4\pi} \int\! d^2 x \sqrt{g}\,T^{ij}\delta
g_{ij} = {\delta\omega\over 2\pi} \int\! d^2 x \sqrt{g}~ T^{ij} g_{ij}  = - {c
\over 24 \pi} \delta \omega \int \! d^2 x\sqrt{g}~ ^{(2)}R =-  {c
\over 3} \delta \omega~.}

This is to be compared with the action \aa\ evaluated on \ab:
\eqn\ai{I = {\Omega_2 \Omega_p \over 16\pi G_{p+3}}\la^3 \ls^p
\int \!d\eta \, \sinh^2 \eta \,{\cal L}_{p+3} + S_{{\rm bndy}}~.}
To make sense of this we need to impose an upper cutoff on $\eta$.
The integral gives
\eqn\aj{\int_0^{\etam} \!d\eta \, \sinh^2 \eta {\cal L}_{p+3}=
\left(-\half \etam + \half \cosh \etam \sinh \etam \right){\cal
L}_{p+3}~.}
Now, $S_{{\rm bndy}}$ is the integral of a expression defined
locally on the AdS boundary.  It is constructed out of the
intrinsic and extrinsic curvature of the boundary.  Such terms
will never give a contribution linear in $\etam$.  Instead, they
cancel the second term in \aj, leaving the action
\eqn\ak{I = -{\Omega_2 \Omega_p \over 32\pi G_{p+3}}\la^3 \ls^p
{\cal L}_{p+3}~ \etam~.}
Comparing \ab\ with \ag\ we have
\eqn\al{ \delta \omega = \delta \etam~,}
which then yields
\eqn\am{c={3 \Omega_2 \Omega_p \over 32\pi G_{p+3}}\la^3 \ls^p
{\cal L}_{p+3}~,}
as we wanted to show.

To summarize, we have shown that the central charge of an AdS$_3
\times S^p \times X$ solution can be obtained simply by
extremizing the central charge function \ac\ with respect to the
AdS and sphere radii.   For a given Lagrangian this just means
solving two algebraic equations.  We will refer to our procedure
as {\it c-extremization}.

We would like to emphasize a couple of important points.  First,
our result applies to an arbitrary higher derivative Lagrangian
including matter fields.  The requirement is just that this
Lagrangian admits a solution with the assumed properties.  Second,
although we used some language familiar from the AdS/CFT
correspondence, our result is completely independent of the
validity of the AdS/CFT conjecture.  Essentially, we have derived
a result about how the gravitational action behaves under Weyl
transformations of the AdS boundary.

\subsec{Black hole entropy}

Once the central charge is known, results for black hole entropy
follow with little additional effort.  We now review how this
works in the general case, allowing independent values of the left
and right moving central charges.

We consider black holes of the form $BTZ \times S^p \times X$. One
way to compute the black hole entropy is by computing the action
of the Euclidean black hole.  From there, one gets the free
energy, and then thermodynamic quantities follow in the standard
way.  The Euclidean BTZ black hole is a solid torus which can be
continued to Lorentzian signature in many different ways. Consider
the cycles on the boundary of the torus which  are noncontractible
with respect to the boundary.  There is clearly one such cycle
which is contractible in the solid torus.  If one calls the
coordinate along the contractible cycle $\phi$, and the other
cycle coordinate $t_E$, then upon continuing $t_E \rightarrow -i
t$ one obtains the geometry of thermal AdS$_3$; that is, global
AdS$_3$ with compact imaginary time.  On the other hand, the
opposite assignment of $t_E$ and $\phi$ leads, upon continuation
to Lorentzian signature, to the  BTZ black hole.\foot{Other
choices lead to the so-called ``$SL(2,Z)$'' family of black
holes.}

From this point of view it becomes clear that the black hole
partition function is just a rewriting of the thermal partition
function.  But the result for the thermal partition function
follows directly from the central charges.  Hence, so too does the
black hole entropy.

Let us illustrate this in more detail; see \refs{\fareytale}.
An asymptotically AdS$_3$ solution carries
energy $H$ and angular momentum $J$.  In the CFT on the boundary
$J$ is the momentum. We can also define the zero modes of the
Virasoro generators as
\eqn\ba{\eqalign{h_L&= L_0-{c\over 24}  = {H-J \over 2}~, \cr h_R &=
\Lt_0-{\ct\over 24} = {H+J \over 2}~.}}
We can think of a bulk solution as a contribution to the partition
function
\eqn\bb{\eqalign{Z(\beta,\mu) =e^{-I}&= \Tr ~e^{-\beta H - \mu J}
\cr & = \Tr ~ e^{ 2\pi i \tau h_L}e^{ -2\pi i \taub h_R}~,}}
where we defined
\eqn\bc{ \tau = i {\beta - \mu \over 2\pi}, \quad \taub = -i
{\beta +\mu\over 2\pi}~.}
When we go to Euclidean signature $\mu$ becomes pure imaginary and
$\taub$ becomes the complex conjugate of $\tau$.  Also, it follows
from \bb\  that $\tau$ is precisely the modular parameter
of the Euclidean boundary torus.

Now consider thermal AdS$_3$.  In Lorentzian signature thermal
AdS$_3$ takes the same form  as AdS$_3$ written in the usual
global coordinates.  On the other hand, we know that global
AdS$_3$ corresponds to the NS-NS vacuum, and as such carries $L_0
= \Lt_0=0$.  Therefore, we conclude that the action of thermal
AdS$_3$ is
\eqn\bd{I_{{\rm thermal}}(\tau,\taub) = {i \pi \over 12}(c\tau -
\ct \taub)~.}

There are in fact additional contributions  due to quantum
fluctuations of massless fields.  \bd\ just takes into account all
the {\it local} contributions.  The extra nonlocal contributions
are, by definition, suppressed for large $\beta$, and will give
subleading contributions to the entropy compared to the local
piece.

We already noted that BTZ is obtained by interchanging $t_E$ and
$\phi$.  This is just a modular transformation of the boundary
torus: $\tau \rightarrow - {1 \over \tau}$.  Recall that a modular
transformation is a diffeomorphism combined with a Weyl
transformation.  The action is invariant since if we take a flat
metric on the torus then all potential anomalies vanish.  We
therefore conclude that
\eqn\be{I_{{\rm BTZ}}(\tau,\taub) = - {i \pi \over 12}({c\over \tau} -
{\ct \over \taub})~.}

From \bb\ it follows that
\eqn\bff{ \eqalign{ h_L & = {-1 \over 2\pi i} {\p I\over
\p\tau}  =  -{ c\over 24 \tau^2}~, \cr  h_R& = {1
\over 2\pi i} {\p I\over \p\taub}  =-{\ct \over 24 \taub^2}~.  }}
From the thermodynamic relation $ I = \beta H + \mu J -S$ we
compute the entropy $S$ to be
\eqn\bg{S_{BTZ}  =  2\pi \left( \sqrt{{c\over 6}h_L}+
\sqrt{{\ct\over 6}h_R}~\right) ~.}

Three facts about this computation are worth stressing.  First,
the result \bg\ holds for an arbitrary theory of gravity admitting
a BTZ black hole (times an arbitrary compactification space).
Second, the result is valid entirely independent of the AdS/CFT
correspondence.  One can just think of it as a result for
computing the action of the Euclidean black hole. Finally, \bg\
gives the entropy in terms of the black hole mass and angular
momentum, and with the central charges appearing as ``undetermined
parameters".  This shows that once we can compute the central
charges, the black hole entropy follows directly from \bg. But we
have seen in the last section how the central charge --- in the
case of $\ct =c$ --- follows from a simple extremization
principle. Altogether, we have arrived at an efficient method
of computing black hole entropy.
  
\subsec{Equivalence with Wald's approach}

The Wald formula \waldlaw\ gives the black hole entropy in an
arbitrary diffeomorphism invariant theory \wald.  In his approach,
one integrates a certain expression over the black hole horizon.
The power of this result is its complete generality. However, for
black holes with a near horizon AdS$_3 \times S^p \times X$
structure, the method we have described above is actually much
simpler to implement.  In particular, one is not required to
locate the horizon at all: c-extremization gives the entropy
directly. In any case, it is worthwhile to check that our result
agrees with Wald's formula, as we do now.

The essential ideas for demonstrating this equivalence appear in
the paper \saidasoda\ where it is shown that Wald's approach leads
to a black hole entropy in the form \bg.  We will follow a
slightly different procedure from \saidasoda.

We first want to write the central charge in a form suitable for
comparison with the Wald formula. It is convenient to work
directly in the theory compactified all the way to  $D=3$. By
assumption, all matter fields take constant values, so that we can
write the action purely in terms of the metric.  In $D=3$ the
Riemann tensor can be expressed in terms of the Ricci tensor; so
the general action will be a function of the Ricci tensor and its
covariant derivatives \foot{ Actually, one can also include a
Chern-Simons term, $S \sim \int \Tr \,\omega\wedge R$, but for now
we exclude such a term. It would lead to $\ct \neq c$ and
associated subtleties, which we postpone till a later section.}
\eqn\bh{ S = {1 \over 16 \pi G_3} \int\! d^3x \sqrt{-g}~
{\cal L}_3 (g^{\mu\nu},R_{\mu\nu})+S_{{\rm bndy}}~. }
Schematically we have
\eqn\bi{{\cal L}_3(g^{\mu\nu},R_{\mu\nu}) \sim \sum_{n} a_n (g^{\mu\nu})^n
(R_{\mu\nu})^n~,}
where the $a_n$ include covariant derivatives and contractions are
not written out explicitly. The central charge function \ac\ is
\eqn\bj{ c(\la) ={3 \Omega_2 \over 32 \pi G_3} \la^3 {\cal L}_3~.}
If we write introduce rescaled variables through \senrescaled\
\eqn\bll{g_{\mu\nu} =\la^2 \hat{g}_{\mu\nu}~, \quad g^{\mu\nu} = {1
\over \la^2} \hat{g}^{\mu\nu}~, \quad R_{\mu\nu} = 2
\hat{g}_{\mu\nu} = \hat{R}_{\mu\nu}~, }
then $\la$ satisfies
\eqn\bk{3{\cal L}_3+ 2 \la^2 {\p {\cal L}_3 \over \p \la^2}=0~.}
Furthermore, in the rescaled variables \bll\ the action reads
\eqn\bm{ S = {1 \over 16 \pi G_3} \int\! d^3x
\sqrt{-\hat{g}}\,\la^3 {\cal L}_3({\hat{g}^{\mu\nu}\over
\la^2},\hat{R}_{\mu\nu})~,}
so the derivative in \bk\ can be evaluated as \foot{We use the
fact that all covariant derivatives vanish on the background.}
\eqn\bn{\la^2 {\p{\cal L}_3 \over \p\la^2} =- \hat{R}_{\mu\nu} {\p {\cal L}_3\over
\p\hat{R}_{\mu\nu}} = - {2 \over \la^2}
g_{\mu\nu} {\p{\cal L}_3\over \p {R}_{\mu\nu}}~~. }
Simplifying \bj\ using \bk\ and \bn\ we find
\eqn\bo{c={\la \over 2G_3} ~g_{\mu\nu} {\p {\cal L}_3 \over \p
{R}_{\mu\nu}}~.}
This formula generalizes the usual Brown-Henneaux central charge
\eqn\bhform{
 c_0= {3\la\over 2G_3}~,
 }
by taking higher derivative corrections into account. The
net result amounts to a rescaling of the $AdS_3$ radius
$\la \to \ell_{\rm eff} =\Omega \la$ where
\eqn\omegadef{
\Omega = {1 \over 3}g_{\mu\nu} {\p {\cal L}_3\over \p
{R}_{\mu\nu}} =
{2 G_3 \over 3\la} c~.}

We are now ready to make the connection with Wald's approach,
since the latter involves an integration over the horizon of the
derivative of the Lagrangian with respect to the curvature. Presently,
the black hole entropy takes the form \bg\ with the central charge \bo.
The BTZ black hole, as usually written, is expressed in terms of the parameters
$M_3$ and $J_3$ which, for a 2-derivative action are identified with
the mass and angular momentum of the black hole. However, in the
presence of higher derivatives the relation is rescaled by the conformal
factor \omegadef\ and we have instead
\eqn\bp{\eqalign{h_{L,R}& = \Omega~ {M_3\mp J_3 \over 2}~. }} We
now find the entropy \bg\
\eqn\br{\eqalign{ S&= {\pi \over 12 G_3}g^{\mu\nu}{\p {\cal L}_3\over \p R_{\mu\nu}}
\left[\sqrt{8G_3\la (M_3+J_3)}+\sqrt{8 G_3 \la (M_3-J_3)}\right] \cr
 &= {A_{\rm BTZ}\over 4G_3} ~\Omega~,
}} where $A_{\rm BTZ}$ is the standard expression for the area of
the BTZ black hole, {\it i.e.} a specific function of $M_3$,
$J_3$, $\la$ and $G_3$; and $\Omega$ is the rescaling factor
\omegadef\ that encodes the  correction due to higher derivative
terms. It is now straightforward to show that Wald's formula
\eqn\bs{\eqalign{S & = -{1 \over 8 G_3}\int_{{\rm hor}} d\phi
\sqrt{g_{\phi\phi}}{ \p {\cal L}_3\over \p R_{\mu
\alpha}}g^{\nu\beta}\epsilon_{\mu\nu}\epsilon_{\alpha\beta}}~,}
agrees precisely with \br, and so with Cardy's formula \bg.


\newsec{Example: M-theory on $CY_3$}

To illustrate our approach, we now consider the example of
M-theory compactified on a Calabi-Yau 3-fold $X=CY_3$, yielding a
supergravity theory in $D=5$.  This is a rich example that includes
includes black holes in both four and five noncompact dimensions
and also BPS black ring solutions.

\subsec{Two-derivative action}

We will follow the conventions in \KLatt, to which we refer for
more details.  In particular, in this section we set $G_5 = {\pi
\over 4}$, which is convenient since it leads to integrally
quantized charges $q^I$.   The hypermultiplets are assumed to be
consistently set to constant values.  Then the $D=5$ action for
the metric and vectormultiplets is given by (as in \aa\ with
$p=2$)
\eqn\ca{\eqalign{ {\cal L}_5  &=  - R + \half G_{IJ} \p_\mu X^I
\p^\mu X^J +  {1 \over 4} G_{IJ} F^{I}_{\mu\nu} F^{J\mu\nu}  +
{\rm (fermions)~} + {\rm ~ (higher~derivs)}\cr S_{CS} &= {1\over
96\pi^2}\int\! d^5 x \,
 C_{IJK}\epsilon_{\mu_1 \ldots \mu_5}
 F^{I\mu_1 \mu_2}
F^{J\mu_3\mu_4}  A^{K\mu_5}   ~.}}
At first we neglect the higher derivative terms.

We consider AdS$_3 \times S^2$ vacua of this theory supported by
magnetic flux. The magnetic charges are given by
\eqn\cb{q^I = - {1 \over 2\pi} \int_{S^2} F^I~ }
where
\eqn\cc{F^I = -{q^I \over 2 \ltwo^2}\epsilon_{S^2}~ }
is interpreted microscopically as $q^I$ M5-branes  wrapped on the
$I$th 4-cycle of $X$. The scalars $X^I$ are taken to have constant
values, fixed by the attractor mechanism to be
\eqn\cd{X^I = {q^I \over ({1 \over 6}C_{IJK}q^I q^J q^K)^{1/3}}~.}
The central charge function \ac\ becomes
\eqn\ce{c(\la,\ltwo)=-6 \la^3 \ltwo^2\left(-{6 \over \la^2}+{2
\over \ltwo^2}-{G_{IJ}q^I q^J \over 4 \ltwo^4}\right)~.}
Extremizing, we find
\eqn\cf{\la =2\ltwo= {1 \over 3}  \sqrt{6 G_{IJ} q^I q^J}, \quad c
= 4\sqrt{2 \over 3} ( G_{IJ} q^I q^J)^{3/2}~.}
Special geometry relations (reviewed in \KLatt) give
\eqn\cg{ G_{IJ}q^I q^J= {3 \over 2} \left({1 \over 6} C_{IJK}q^I
q^J q^K \right)^{2/3}~, }
which then yields
\eqn\ch{\la=2\ltwo= \left({1 \over 6} C_{IJK}q^I q^J q^K
\right)^{1/3} ,\quad  c= C_{IJK} q^I q^J q^K~.}
In an appendix we review how these relations appear  in the
explicit solutions.

\subsec{Higher derivative corrections}
Our approach makes it simple to include the effects of higher derivatives.
As an example we consider adding to the action the term
\eqn\ci{ \Delta {\cal L}_5 = A
\left(R^{\mu\nu\alpha\beta}R_{\mu\nu\alpha\beta}-4R^{\mu\nu}
R_{\mu\nu}+R^2\right)~,}
for some constant $A$.  If we were in  $D=4$ this term would be
the Euler invariant. It is one particular higher derivative term
present in M-theory on $CY_3$.  Since other terms are present as
well, we don't expect \ci\ to capture the complete microscopic
correction to the central charge or the black hole entropy. Later,
we will do better, but this example is a good illustration.

Evaluated on AdS$_3 \times S^2$ we have
\eqn\cj{ \Delta {\cal L}_5 = -{24 A \over \la^2 \ltwo^2}~,}
and so the central charge function is now
\eqn\ck{c(\la,\ltwo)=-6 \la^3 \ltwo^2\left(-{6 \over \la^2}+{2
\over \ltwo^2}-{G_{IJ}q^I q^J \over 4 \ltwo^4}-{24 A \over \la^2
\ltwo^2}\right)~.}
Extremizing and using \cg, we find that both the radii and the
central charge are corrected:
\eqn\cl{\eqalign{\la& = \left({1 \over 6}C_{IJK}q^I q^J
q^K\right)^{1/3} + {4 A \over \left({1 \over 6}C_{IJK}q^I q^J
q^K\right)^{1/3}} + O(A^2) \cr \ltwo& =\half \left({1 \over
6}C_{IJK}q^I q^J q^K\right)^{1/3} + { A \over \left({1 \over
6}C_{IJK}q^I q^J q^K\right)^{1/3}} + O(A^2) \cr c& =C_{IJK}q^I q^J
q^K + 144 A\left({1 \over 6}C_{IJK}q^I q^J q^K\right)^{1/3}
+O(A^2) ~.}}

\subsec{Black hole entropy} The corrected central charge \cl\
gives the black  hole entropy according to \bg\
\eqn\cm{ S= 2\pi \sqrt{{c\over 6}h_L}+2\pi
\sqrt{{c\over 6}h_R}~.}
This formula could refer to either asymptotically AdS or
asymptotically flat black holes, with slightly different
interpretations.  In the AdS case, the Virasoro generators are
related to the mass and angular momentum of the black hole
as in \ba.
The entropy formula \cm\ then holds for an arbitrary ({\it i.e.}
nonextremal, nonsupersymmetric)  BTZ black hole in this theory.

\lref\Emparan{
  R.~Emparan and D.~Mateos,
  ``Oscillator level for black holes and black rings,''
  arXiv:hep-th/0506110.
}

\lref\LarsenUK{
  F.~Larsen and E.~J.~Martinec,
  ``U(1) charges and moduli in the D1-D5 system,''
  JHEP {\bf 9906}, 019 (1999)
  [arXiv:hep-th/9905064];
  ``Currents and moduli in the (4,0) theory,''
  JHEP {\bf 9911}, 002 (1999)
  [arXiv:hep-th/9909088].
}

 In the asymptotically flat case \cm\ still holds, but additional
work is required to relate $h_L$ and $h_R$ to the asymptotic
charges of the black hole.   A good example is the case of
M5-branes and M2-branes in M-theory on $CY_3\times S^1$.  This was
the case considered in \MSW.  As before, we consider the M5-branes
(with charges $q^I$) to wrap  4-cycles in $CY_3$, and in addition
we take the M2-branes (with charges $Q_I$) to wrap 2-cycles. The
asymptotically flat solution, in the case of $CY_3=T^6$
compactification, was given in \iosef. After taking the near
horizon limit (which was all that was needed for the analysis in
\MSW) we find that AdS$_3$ becomes a extremal rotating BTZ black
hole, with
\eqn\cn{ h_R = Q_0 +{1\over 2}C^{IJ}Q_I Q_J~.}
Here $Q_0$ is momentum running around the asymptotic $S^1$, {\it
i.e.} the Kaluza-Klein electric charge, and $C^{IJ}$ is related to
the intersection matrix of the compactification manifold. The
extra term in $h_R$ is due to the nonzero M2-brane charges. More
discussion of this effect can be found in
\refs{\MSW,\LarsenUK,\Emparan}. With this identification \cm\
gives the entropy in terms of the charges measured at asymptotic
infinity. As we have already stressed, once higher derivatives are
included \cm\ will  still hold but the central charges will be
corrected. 

\newsec{Anomalies, central charges, and entropy}
Up until this point we have restricted attention to cases with
$\ct =c$, and focussed on computing the central charge and black
hole entropy from the conformal anomaly.  The approach is quite
powerful, but for certain cases one can do even better.  Indeed,
one potential disadvantage is that to compute the conformal
anomaly one needs to know all the terms in the action which are
nonzero in the given background. But all such terms are not
necessarily known when one is considering higher derivative
theories.   Furthermore, when $\ct\neq c$, this approach is
clearly insufficient to determine both central charges.

It is wise to take advantage of any other anomalies in the
problem, as well as the relations among them following from
symmetries. For the M5-brane example considered in the previous
section gravitational anomalies are especially powerful.  As we
will review, there are two anomalies --- the tangent and normal
bundle anomalies --- which follow from knowledge of a single term
in the action, and which suffice to determine the corrections to
both central charges \refs{\HMM}. So from this point of view
the corrected entropy formula for the M5-brane
emerges rather easily.

In the absence of gravitational anomalies the on-shell bulk
supergravity action is a diffeomorphism invariant function of the
boundary geometry.  By the AdS/CFT correspondence it is supposed
to yield the partition function of the CFT on the boundary.  In
the presence of gravitational anomalies, one is still led to
conjecture the correspondence, but with each side suffering a loss
of diffeomorphism invariance.  This manifests itself in the
non-conservation of the boundary stress tensor.

\subsec{Some higher derivative terms}
\lref\HoweCY{
  P.~S.~Howe and D.~Tsimpis,
  JHEP {\bf 0309}, 038 (2003)
  [arXiv:hep-th/0305129].
}

Several higher derivative terms in the effective action of
M-theory are known (some relevant references are \refs{\generalRRRR,\tseytRRRR,\HoweCY}).
Those involving $R^4$ terms take the schematic forms
 \eqn\dz{\eqalign{ & t_8 t_8 RRRR~, \cr
& \epsilon_{11}\cdot\epsilon_{11} RRRR~,  \cr
& \epsilon_{11} C_3 \left[\Tr R^4-{1 \over 4}(\Tr R^2)^2\right]~.}}
For the precise definitions of these, and their coefficients in
the action, see, {\it e.g.} \tseytRRRR. Of most interest to us is
the term given in the third line since this term yields
corrections to central charges and black hole entropy. The
coefficient of this term is determined by requiring that its
anomalous variation under diffeomorphisms cancel anomalous terms
on the M5-brane worldvolume. We will review this in the
dimensionally reduced context below.

Dimensional  reduction of these terms on $CY_3$ leads to various
higher derivative terms in $D=5$ \antRRRR, as well as
shifts in the coefficients of some two-derivative terms.  One of the terms
that appear this way is the dimensionally
continued Euler invariant \ci. Here we focus on
\eqn\da{S_{\rm anom} = {c_2 \cdot P_0 \over 48} \int_{M_5} A
\wedge p_1~,}
which arises from reduction of the third term in \dz. In \da\
$p_1$ is the first Pontryagin class
\eqn\db{ p_1 = - \half \left({1\over 2\pi}\right)^2 \Tr R \wedge
R~.}
We take the M5-brane to wrap the cycle $P_0 = P_0^I \sigma_I$,
where $\{\sigma_I\}$ form a basis for $H_4(X,Z)$.   The choice of
4-cycle then determines a particular linear combination of gauge
fields in five dimensions, which was denoted by $A$ in \da. Finally, $c_2$
is the second Chern class of $X$, which has coefficients $c_{2I}$ in its
expansion with respect to chosen basis for $H^4(X,Z)$.

After reduction on $X$, the wrapped M5-branes correspond to a
string in five dimensions, on which lives a chiral CFT.  As
explained in \HMM, the term \da\ cancels the gravitational
anomalies of the CFT.\foot{More precisely, it cancels the part of
the anomaly linear in M5-brane charge.  There are also cubic
terms which we'll discuss momentarily.}

\subsec{Anomalies}

Anomaly cancellation occurs via the inflow mechanism, as we now
recall.  First of all, since $A$ is ill-defined in the presence of
a magnetic charge, \da\ should really be written after performing
an integration by parts and discarding the boundary term.  So the
actual term of interest is
\eqn\dc{S = \half \left({1\over 2\pi}\right)^2{c_2 \cdot P_0 \over
48} \int_{M_5} F \wedge \omega_3~,}
where $\omega_3$ is the Lorentz Chern-Simons 3-form:
\eqn\dd{ \omega_3 = \Tr (\omega d\omega +{2 \over 3}\omega^3)~,}
with $\omega$ being the spin connection.   Now under a local
Lorentz transformation parameterized by $\Theta$,
\eqn\de{\delta \omega = d\Theta + [\omega,\Theta]~,}
the action changes as
\eqn\df{\delta S_{\rm bulk} = \half \left({1\over
2\pi}\right)^2{c_2 \cdot P_0 \over 48} \int_{M_5} F \wedge \Tr
(d\Theta\wedge d\omega)~.}

At this point we encounter two distinct interpretations. The
approach of \HMM\ was to consider the magnetic string essentially
as a pointlike  defect placed in an ambient space. The presence of
the magnetic string corresponds to $dF$ having delta function
support at the location of the string.  In this approach, one
integrates \df\ by parts, and then uses the delta function to
perform the integral over the directions transverse to the string.
What remains is an integral over the string worldvolume, which
cancels a term coming from the variation of the path integral over
the string degrees of freedom.

The interpretation in our case is somewhat different.  We are
dealing with a smooth supergravity solution with geometry
AdS$_3 \times S^2\times X$ and $dF=0$.   The branes have been replaced by flux.
Instead of cancelling the anomaly at the brane location, we get a
contribution at the AdS boundary.  It is clear that this
contribution yields the anomalous variation of the CFT on the
boundary.  This mechanism is well known in AdS/CFT, going back to
the treatment of the R-symmetry anomaly of ${\cal N}=4$
super-Yang-Mills in \wittenAdS.  In particular,  \df\ gives the
boundary term
\eqn\dg{\delta S_{\rm bulk} = \half ~ \left({1\over
2\pi}\right)^2~ {c_2 \cdot P_0 \over 48} \int_{\p M_5} F \wedge \Tr
(\Theta d\omega)~.}
We consider pure AdS$_3 \times S^2$ with the components of
$F$ given by \cc. Integrating \dg\ over the $S^2$ we obtain
\eqn\dh{\delta S_{\rm bulk } =  - \half  ~{c_2 \cdot q \over 48}~
{1\over 2\pi} \int_{\p {\rm AdS}_3}  \Tr (\Theta d\omega)~.}
Importantly the matrices $\Theta$  and $d\omega$ are still by
$5\times 5$; they include indices along the $AdS_3$ boundary and
also in the radial and $S^2$ directions. Accordingly, we can study
two kinds of anomalies, associated with diffeomorphisms that map
the boundary to itself (tangent bundle anomaly) and with
diffeomorphisms acting on the vectors normal to the boundary
(normal bundle anomaly). From the point of view of the D=2 CFT,
these are gravitational and $SU(2)_R$ symmetry anomalies.

In the CFT the gravitational anomaly is obtained
via descent from $I_4 = 2\pi ~{1 \over 24} (c-\ct)p_1$, yielding
\eqn\dha{\delta S_{\rm CFT } =  {c- \ct  \over 48} ~{1\over 2\pi}
\int_{\p {\rm AdS}_3}  \Tr (\Theta d\omega)~.}
Equating this with \dh\ we find \foot{There are two (cancelling)
sign changes relative to the anomaly inflow in \HMM: the boundary
at infinity has normal opposite to that of a defect in bulk; and
also we are {\it equating} the two anomalies, as in AdS/CFT,
rather than {\it cancelling}  them, as in the anomaly inflow.}
\eqn\di{\ct- c = \half c_2 \cdot q~.}
The computation of the normal bundle anomaly is similar. In this
case the corresponding CFT anomaly is in the $SU(2)_R$ symmetry
which, in our conventions, acts on the leftmovers so that the
normal bundle anomaly contributes
\eqn\dj{
c_{\rm lin} = \half c_2 \cdot q~,
}
to $c$. The form of \di\ and \dj\ are the same because these
contributions arise from the same anomaly \dh, decomposed into
tangent and normal bundle part, and interpreted appropriately.
These expressions capture the linear contributions to
the central charges exactly. However, there are also ${\cal O}(q^3)$
contributions (see \ch) coming from the
two-derivative part of the action, and so altogether we have
\eqn\dk{c = C_{IJK}q^I q^J q^K + {1 \over 2} c_2 \cdot q~, \quad
\ct = C_{IJK}q^I q^J q^K + c_2 \cdot q~.}
These are the results found in \refs{\MSW,\HMM}.

\lref\Mfiveano{  D.~Freed, J.~A.~Harvey, R.~Minasian and G.~W.~Moore,
  ``Gravitational anomaly cancellation for M-theory fivebranes,''
  Adv.\ Theor.\ Math.\ Phys.\  {\bf 2}, 601 (1998)
  [arXiv:hep-th/9803205].
}
The $C_{IJK}q^I q^J q^K$ contributions are actually quite subtle in
the context of anomaly cancellation for  M5-branes viewed as
pointlike defects \refs{\WittenMfive,\Mfiveano}. The ${\cal O}(q^3)$
contribution to the normal bundle anomaly requires a subtle
modification of the M-theory Chern-Simons term.  By contrast, in
the context of the smooth supergravity backgrounds considered here,
this contribution is simple to understand because it comes from the
leading two-derivative part of the action.  We have phrased this
in terms of computing the conformal anomaly, but we could have
equally well computed the normal bundle anomaly directly.  In our
problem supersymmetry related these anomalies to one another,
so a computation of either suffices.

\subsec{Application: heterotic strings}

An important special case of our computations is M-theory  on
$K3\times T^2$. Consider an $M5$-brane wrapped around the $K3$ and
transverse to the $T^2$. In this case we have $C_{IJK}q^I q^J
q^K=0$, and $c_2\cdot q =24$ because the Euler number of $K3$ is
$24$. Therefore, \dk\ gives $c=12$ and $\ct=24$. These are the
correct assignments for the heterotic string which,
indeed, is a dual representation of an M5-brane on $K3\times T^2$.
Thus we find the central charges of {\it both} sides of the
heterotic strings; so we are sensitive to all excitations, rather
than just the BPS states. In particular, from the Cardy formula
\hetex\ we get the entropy of nonsupersymmetric small black holes
in agreement with the non-BPS entropy of the heterotic string.
Although the formulae \dk\ have been known for some time, this
agreement apparently has not been noticed before.

\lref\KutasovZH{
  D.~Kutasov, F.~Larsen and R.~G.~Leigh,
  ``String theory in magnetic monopole backgrounds,''
  Nucl.\ Phys.\ B {\bf 550}, 183 (1999)
  [arXiv:hep-th/9812027].
} The recovery of both the central charges of the heterotic string
sounds like an extremely powerful and surprising result when put,
as above, in terms of  the near horizon geometry, corrected by
higher derivative terms in the action. However, from another point
of view the agreement is almost trivial: a heterotic
string propagating in a curved background suffers gravitational
anomalies, because $c\neq \ct$, and these must be cancelled by
bulk terms, {\it via} the inflow mechanism. This works, of course;
indeed, it would be one way to derive the anomalous coupling
$S_{\rm anom}$, including the coefficient. Related to this,
heterotic string theory in $AdS_3\times N$ has linear corrections
that precisely reproduce the ones seen here \KutasovZH. From
either point of view, we should hardly be surprised when these
couplings give back the heterotic string, when interpreted in
terms of the near horizon geometry and its boundary at infinity.
On the other hand, the fact that the agreement is essentially
automatic does not make it any less valid, nor any less
interesting.

\subsec{Application: inclusion of angular momentum}

Consider the BPS states of a heterotic string wrapped on an $S^1$
in $T^5$, with fixed winding number, rightmoving momentum, and
angular momentum in a given 2-plane.  The microscopic entropy is
known to be \russuss\
\eqn\yz{ S=4\pi \sqrt{N_w N_p - J}~.}
Geometrically, the states correspond to rotating helical strings.
The maximal angular momentum, $J= N_w N_p$ is attained when all
the momentum is placed in oscillators of the lowest mode number,
with polarizations in the angular momentum 2-plane.  The profile
of the helix is then a circle.  If we decrease $J$ from its
maximal value while holding $N_{w,p}$ fixed, then there are
additional microstates in which the string wiggles away from its
circular shape, either in the noncompact or internal dimensions.
These additional states give rise to the entropy \yz. As we will
now argue, there is also a black object with the same charges and
whose entropy agrees with \yz.

Using heterotic/IIA duality, and lifting to M-theory, the
configuration above describes a rotating helical M5-brane wrapped
on $K3 \times T^2$.  The supergravity solution will have near
horizon limit AdS$_3 \times S^2 \times K^3 \times T^2$.   The
rightmoving central charge is $\tilde{c} = 24 N_w$, since the
M5-brane wraps $K3$ $N_w$ times.  The level number $h_R$ appearing
in the near horizon region differs from the total rightmoving
momentum $N_p$ measured at infinity.  Although we have not checked
this explicitly in the present context, in other very similar
cases (see, e.g. \MarolfFY) one finds that $h_R$ is obtained by
subtracting from $N_p$ the momentum used up by the gyration:
\eqn\yy{ h_R = N_p  -N_{{\rm gyro}}~.}
The mechanical gyration of the string carries momentum and angular
momentum related by $J_{{\rm gyro}} = {\lambda \over 2\pi} P_{{\rm
gyro}}$, where $\lambda$ is the wavelength of the gyration.  Since
our brane is wrapped $N_w$ times around a circle of radius $R$,
the largest possible wavelength (which yields the highest entropy)
is $\lambda = 2\pi R N_w$, and so
\eqn\yx{ h_R = N_p -{J \over N_w}~.}
The near horizon geometry will thus be a BTZ black hole with
entropy given by the Cardy formula as
\eqn\yw{S = 2\pi \sqrt{ {c\over 6} h_R} = 4\pi \sqrt{N_w N_p -
J}~,}
in agreement with \yz.  The black object could be thought of as a
``small" black ring.

With the replacements $N_w \rightarrow N_5$ and $N_p \rightarrow
N_1$, \yz\ also gives the ground state entropy of the D1-D5 system
on K3.  Indeed there is a duality chain that relates the two
systems.  The M-theory configuration can be interpreted as IIA on
$K3 \times S^1$ with NS5-branes wrapped on the compact space and
carrying momentum on the $S^1$.   A T-duality on the $S^1$
followed by S-duality then yields the D1-D5 system. The ground
state entropy of the D1-D5 system has recently been obtained in a
different approach by Iizuka and Shigemori \masakiref.

\newsec{Discussion: corrections to all orders in $1/Q$ and beyond}
The black hole entropy discussed in this paper has been presented
in all cases in terms of the Cardy formula which is essentially
semi-classical. It is interesting to think about how further
corrections might be included. In particular, recent work has
shown that is possible to reproduce the BPS entropy of the
heterotic string to all orders in an expansion in inverse powers
of the charges \DDMP. Let us now show how our approach is
naturally extended to include this agreement.

In evaluating the black hole partition function in section 2.3 we
specified the black hole temperature $\beta$ and chemical
potential $\mu$, which are conjugate to the mass and angular
momentum of the black hole.  We now note that we could also
specify the values of any conserved charges or, alternatively,
the boundary values of the corresponding gauge potentials.

In the case of M-theory on $K3 \times T^2$ we have gauge fields
$A^I$ that couple to charges  $Q_I$ that correspond to wrapped
M2-branes.   We thus need the Euclidean action of black holes
carrying these charges (as well as the M5-brane charge $q^I$).
According to \cn\ this just gives a shift in $h_R$ which, from \bb,
changes the action to
\eqn\xc{I_{BH}(\overline{\tau}, Q_I) = {i\pi \over 12}  {\ct \over \overline{\tau} }
+2\pi i \taub~ {1
\over 2} C^{IJ}Q_I Q_J~.}
To focus on BPS states  we set the left moving temperature to
zero: ${1 \over \tau} =0$. Semi-classically, the potentials are
related to the charges as \eqn\potcomp{ \phi^I = {1\over\pi}
{\partial I_{BH}\over\partial Q_I}~, } so
\eqn\xi{ I_{BH}(\phi^0,\phi^I) =
{\pi \ct \over 6 \phi^0} -{ \pi
\over 2} {C_{IJ}\phi^I \phi^J\over \phi^0}~,}
where we renamed the right moving temperature
\eqn\xf{\phi^0 = {2\over i} \taub~.}
The potentials $\phi^0,\phi^I$ defined in \potcomp\ and \xf\ were
designed to agree with the conventions in the topological string
literature \refs{\OSV,\DDMP} which amounts to the equality
\eqn\xh{ e^{-\pi( Q_0 \phi^0 + q_I \phi^I)} = e^{2\pi i\taub (Q_0
+ Q_I A^I_t)}~. } Now, the expression \xi\ for the action is
precisely the same as (the negative of) the free energy ${\cal
F}_{\rm pert}$ appearing in (2.6) of \DDMP, and at this point we
can simply follow their analysis.  In particular, the degeneracy
of states $\Omega(Q_0,Q_I)$ with the specified charge is given by
the relation between the canonical and microcanonical ensembles
\eqn\xf{ \Omega(Q_0,Q_I) = \int\! d\phi\, e^{-I_{BH}(\phi^0,\phi^I) +
\pi(Q_0 \phi^0 + Q_I \phi^I)}~.}
Carrying out the integral yields a Bessel function which correctly
accounts for the number of heterotic string states to all orders in
inverse powers of charges.  We refer the reader to \DDMP\ for
the details (and also to \senrescaled\ for an alternative approach).
The point we wish to emphasize here is that the power law
corrections to the black hole entropy are semi-classical
in nature, and so they can be captured by our approach.

Ultimately, several other corrections must be included in order to
account completely for the microscopic degeneracies including
exponentially suppressed terms. For example, there are
contributions from world-sheet and brane  instantons and also,
more dramatically, from semi-classical geometries distinct from
the one contributing to the leading term.  These corrections
remain to be understood, both in the 4D topological string
approach, and in the approach considered here.

\bigskip
\noindent {\bf Acknowledgments:} \medskip \noindent We would like
to thank Masaki Shigemori for discussions and an advance copy of \masakiref.
We thank the organizers of the Amsterdam
workshop on Strings and Branes for hospitality as this work was
completed. Also thanks to P. van Hove for correspondence.
The work of PK is supported in part by the NSF grant
PHY-00-99590. The work of FL is supported in part
by the DoE.

\appendix{A}{Asymptotically flat M5-brane solution}

For convenience, we give here the asymptotically flat solution
representing M5-branes wrapped on 4-cycles of $CY_3$.  We follow
the conventions of \KLatt. The $CY_3$ has harmonic $(1,1)$ forms
$J_I$ and Kahler moduli $X^I$. The metric and 3-form are
\eqn\ja{\eqalign{ds^2& = ds_5^2 +ds_{CY_3}^2\cr {\cal A}& = A^I
\wedge J_I~}}
with
\eqn\jb{\eqalign{ ds_5^2&=({1 \over 6}C_{IJK}H^I H^J
H^K)^{-1/3}(-dt^2+dx_4^2)+({1 \over 6}C_{IJK}H^I H^J
H^K)^{2/3}(dr^2+r^2 d\Omega_2^2) \cr A^I &= \half q^I (1+\cos
\theta)d\phi \cr X^I & = {H^I \over ({1 \over 6}C_{IJK}H^I H^J
H^K)^{1/3}}\cr H^I& =\Xb^I+ {q^I \over 2r}~.}}
To examine the near horizon geometry we write
\eqn\jba{r={ {1 \over 6}C_{IJK}q^I q^J q^K \over 2 z^2}~.}
For $z\rightarrow \infty$ we then find the following AdS$_3 \times
S^2 \times CY_3$ geometry
\eqn\jbb{\eqalign{ds_5^2 & =  \la^2 {-dt^2 + dx_4^2 +dz^2 \over
z^2} + \ltwo^2 d\Omega_2^2 \cr X^I &= {q^I \over ({1 \over
6}C_{IJK}q^I q^J q^K)^{1/3} }    }}
with
\eqn\jbc{ \la = 2\ltwo =\left({1 \over 6}C_{IJK}q^I q^J
q^K\right)^{1/3}~.}
The Brown-Henneaux computation of the central charge applied to
this case gives
\eqn\jbd{ c= C_{IJK}q^I q^J q^K~.}
\jbc\ and \jbd\ are in perfect agreement with \ch.

\listrefs

\end

\subsec{Application: D1-D5 system with nonzero angular momentum}
An interesting generalization is to seek a black hole
representation of {\it rotating} heterotic strings. After duality
we can think of this as the D1-D5 system with non-zero angular
momentum (equivalently, R-charge in the CFT). Wrapping the D5 on
K3 the entropy becomes \russuss\
\eqn\ya{S= 4\pi \sqrt{N_1 N_5 - J_\psi}~,}
where $J_\psi$ is the angular momentum in a transverse 2-plane. We
can see how this entropy formula emerges in our approach using
ideas from a recent analysis by Iizuka and Shigemori \masakiref.

The presence of nonzero $J_\psi$ but vanishing $J_\phi$ means that
the corresponding supergravity solution is a 2-charge supertube
\supertube: D1-D5$\rightarrow$ kk, {\it i.e.} the D1-D5 is
dissolved in a Kaluza-Klein monopole which is wrapped into a
topologically trivial ring. Such solutions exist for integer
winding number for the  kk-monopole, with $1 \leq n_{kk} \leq N_1
N_5$. Now dualize this to the M-theory frame. The D1 and D5 turn
into M2-branes and the kk-monopole into an M5-brane.  This set of
charges will flow in the IR to AdS$_3 \times S^2 \times K3 \times
T^2$.  The central charge is  fixed by the M5-brane as
\eqn\yb{ \tilde{c} = c_2 \cdot p = 24 n_{kk}~. }

The Cardy formula gives us the entropy in terms of $c$ and
$h_{L,R}$.    As in \MSW\ we have\foot{Use $D_{IJK} = {1 \over
6}C_{IJK},~ D_{IJ} =D_{IJK}q^K,~ D^{IJ}D_{JK}=\delta^I_K$.}
\eqn\yc{ h_R = -J_\psi +{1 \over 12}D^{IJ}Q_I Q_J = -J_\psi + {N_1
N_5\over n_{kk}}}
and $h_L=0$.   This gives the entropy
\eqn\yd{ S =4\pi \sqrt{N_1 N_5 - n_{kk}J_\psi}~.}
The highest entropy occurs for $n_{kk}=1$, which yields \ya. This
entropy represents the entropy of a small black ring \masakiref;
i.e. a ``small" version of the black rings found in \refs{\rings}

When the angular momentum bound is saturated, $J_\psi = N_1 N_5$,
the entropy vanishes and the supergravity solution becomes a
smooth, horizon free, supertube solution.  Taking slightly less
$J_\psi$ would yield a  singular ring solution in the two
derivative theory.  Here we see that higher derivatives yield an
event horizon, and the correct microscopic entropy.